\documentclass[aps,pra,twocolumn,showpacs,amsmath,amssymb,preprintnumbers,superscriptaddress,10pt]{revtex4-2}
\usepackage{graphicx}
\usepackage{subfig}
\graphicspath{{figures/}}
\usepackage{siunitx}
\usepackage{hyphenat}
\usepackage{braket} 
\usepackage[bookmarks=false]{hyperref}
\usepackage{color}
\usepackage{verbatim}
\usepackage[capitalise]{cleveref}
\crefname{section}{Sec.}{Secs.}
\Crefname{section}{Section}{Sections}
\usepackage{caption}
\usepackage{graphicx}
\usepackage{float} 
\usepackage{CJKutf8}
\usepackage{ragged2e}
\usepackage{diagbox}
\usepackage{booktabs}
\usepackage{array}
\usepackage{tabularx}
\definecolor{pink}{RGB}{255,0,255}

\definecolor{red}{rgb}{0,0,1}

\begin{document}
\newcolumntype{P}[1]{>{\centering \arraybackslash}p{#1}}
\newcolumntype{L}{X}
\newcolumntype{C}{>{\centering \arraybackslash}X}
\newcolumntype{R}{>{\raggedright \arraybackslash}X}

\title{Intensity correlations in measurement-device-independent quantum key distribution}

\author{Junxuan Liu}
\thanks{These authors contributed equally}
\affiliation{Institute for Quantum Information \& State Key Laboratory of High Performance Computing, College of Computer Science and Technology, National University of Defense Technology, Changsha, 410073, China}

\author{Tianyi Xing}
\thanks{These authors contributed equally}
\affiliation{Institute for Quantum Information \& State Key Laboratory of High Performance Computing, College of Computer Science and Technology, National University of Defense Technology, Changsha, 410073, China}

\author{Ruiyin Liu}
\affiliation{Institute for Quantum Information \& State Key Laboratory of High Performance Computing, College of Computer Science and Technology, National University of Defense Technology, Changsha, 410073, China}

\author{Zihao Chen}
\affiliation{Institute for Quantum Information \& State Key Laboratory of High Performance Computing, College of Computer Science and Technology, National University of Defense Technology, Changsha, 410073, China}

\author{Hao Tan}
\affiliation{China Telecom Quantum Information Technology Group Co., Ltd, Hefei 230088, China}

\author{Anqi Huang}
\email{angelhuang.hn@gmail.com}
\affiliation{Institute for Quantum Information \& State Key Laboratory of High Performance Computing, College of Computer Science and Technology, National University of Defense Technology, Changsha, 410073, China}

\date{\today}

\begin{abstract}
The intensity correlations due to imperfect modulation during the quantum-state preparation in a measurement-device-independent quantum key distribution (MDI QKD) system compromise its security performance. Therefore, it is crucial to assess the impact of intensity correlations on the practical security of MDI QKD systems. In this work, we propose a theoretical model that quantitatively analyzes the secure key rate of MDI QKD systems under intensity correlations. Furthermore, we apply the theoretical model to a practical MDI QKD system with measured intensity correlations, which shows that the system struggles to generate keys efficiently under this model. We also explore the boundary conditions of intensity correlations to generate secret keys. This study extends the security analysis of intensity correlations to MDI QKD protocols,  providing a methodology to evaluate the practical security of MDI QKD systems.
\end{abstract}
\maketitle

\section{Introduction}
\label{sec:intro}
Quantum key distribution (QKD), based on the laws of quantum physics, promises information-theoretic security of sharing secret key between two remote parties, Alice and Bob~\cite{bennett1984,PhysRevLett.68.3121}. However, in practice, the characteristics of physical devices cannot be fully matched to the theoretical models. This discrepancy allows eavesdroppers to exploit imperfections in physical devices to compromise the security of QKD systems~\cite{PhysRevA.74.022313,Lamas-Linares:07,Lydersen2010,Lydersen:10,Xu_2010,PhysRevA.84.062308,Wiechers_2011,Gerhardt2011,PhysRevA.83.062331,PhysRevLett.107.110501,PhysRevLett.112.070503,PhysRevA.91.062301,PhysRevA.92.022304,7571108,PhysRevA.94.030302,PhysRevA.98.012330,PhysRevApplied.10.064062,PhysRevApplied.12.064043,PhysRevApplied.13.034017,PRXQuantum.3.040307,e24020260,Chaiwongkhot2022,PhysRevApplied.19.014048}.

To protect QKD systems from eavesdropping attacks, researchers have made significant contributions. For example, measurement-device-independent~(MDI) QKD was proposed, removing the security assumptions associated with the measurement and ensuring that the security of the QKD system is no longer threatened by potential vulnerabilities in the measurement unit~\cite{PhysRevLett.108.130503}. Additionally, the decoy-state protocol was proposed to enhance the security performance and efficiency of a QKD system with phase-randomized weak coherent sources~\cite{PhysRevLett.94.230503,PhysRevLett.94.230504,PhysRevA.72.012326}. While enhancing the security of QKD systems, efforts have also been made to improve the key rate of the system by raising the repetition rate of key distribution. The exploration of high-speed QKD systems has become a prominent area of research.

In high-speed QKD systems, either running BB84 QKD or MDI QKD protocols, some vulnerabilities in the transmitters may become more noteworthy and further threaten the practical security of QKD implementation, such as the correlation in the intensity modulation. Due to the memory effect of optical modulators and the cross-talk of their electrical drivers, the intensity modulated by the transmitter's intensity modulator may be influenced by previous states, leading to correlation between the intensities of adjacent pulses~\cite{Yoshino2018_11,Roberts:18,Lu2021,9907821}. This correlation violates the principle of independence between quantum states and compromises the security of the QKD system.

To address the issue of correlation in the transmitter, one approach is to model the QKD system under correlation, and perform a quantitative analysis on the impact of correlation on the security of the QKD system. Reference~\cite{PereiraSciAdv} introduced the reference technique (RT), quantifying the state preparation flaws with correlations. Subsequently, Refs.~\cite{Zapatero2021securityofquantum,PhysRevApplied.18.044069} utilized the Cauchy-Schwarz (CS) inequality to estimate the impact of intensity correlations in BB84 QKD systems. However, the imperfection of intensity correlations has not been considered in the security analysis models of MDI QKD protocol. Thus, the impact of intensity correlations in a practical MDI QKD system has not been quantified.

In this work, we present a theoretical model to analyze the impact of the intensity correlations on the practical security of a MDI QKD system. This model characterizes the effect of the intensity correlations on the security of the MDI QKD by describing the discrepancy between the actual intensity under correlation and the ideal intensity setting. According to the proposed security model, we simulate the secret key rate with the different relative discrepancy and correlation length. Furthermore, the security model is applied to analyze the security performance of a practical MDI QKD system with quantified intensity correlations measured in experiment.\par

This study proposes the security model to analysis the intensity correlations and shows its feasibility by applying to a practical MDI QKD system. Simulations of the key rate reveal that intensity correlations significantly compromise the security of the practical MDI QKD system, making it an unavoidable side channel. We also provide boundary conditions of intensity correlations to achieve the secure key rate. Thus, this study offers references and recommendations for the MDI QKD implementation to limit its intensity correlations.

The paper is structured as follows. In~\cref{sec:assumptions}, we define necessary parameters and clarify security assumptions, as foundation for the subsequent security proof. In~\cref{sec:PROOF}, we present the security model of MDI QKD by considering the intensity correlations in the yield and error probability. In~\cref{sec:Simulations}, we simulate the results of the secret key rate under intensity correlations in a MDI QKD system and analyze the impact of two correlation parameters\textemdash correlation length and relative deviation\textemdash on its security performance. The theoretical model is used to assess the impact of intensity correlations on the security of a practical MDI QKD system in~\cref{sec:Application}, and this study is concluded in~\cref{sec:Conclusion}.

\section{Parameter definitions and security assumptions}
\label{sec:assumptions}
To simplify, we consider a symmetric MDI QKD protocol with three-intensity decoy-state and polarization-encoding scheme. It is noted that this analysis can be extended to the asymmetric MDI QKD protocol. Additionally, this framework of security proof is applicable to different numbers of intensity setting and other encoding schemes.

In each round $k$ $(k=1,2,...,N)$ of protocol, Alice (Bob) selects an intensity setting $a_k$ ($b_k$) $\in A=\{\mu,\nu,\omega\}$ with probability $p_{a}$ ($p_{b}$), a basis $x_k^A$ ($x_k^B$) $\in Basis=\{X,Z\}$ with probability $q_{x}^A$ ($q_{x}^B$), and a uniform raw key bit $r_k^A$ ($r_k^B$) $\in \mathbb{Z}_2=\{0,1\}$. Without loss of generality, we assume that the intensity settings satisfy the condition $\mu > \nu > \omega \geq 0 $. Alice (Bob) encodes one of the four states defined by $x_k^A$ ($x_k^B$) and $r_k^A$ ($r_k^B$) in a phase-randomized weak coherent source with intensity setting $a_k$ ($b_k$), then sending it to Charlie through the quantum channel. However, the actual mean photon number of each pulse emitted by Alice (Bob) may not correspond to her (his) intensity setting $a_k$ ($b_k$) due to intensity correlations, denoted as $\alpha_k$ ($\beta_k$). Let $\vec{a}_k = a_k,a_{k-1},...,a_1$ ($\vec{b}_k = b_k,b_{k-1},...,b_1$) represent the sequence of Alice's (Bob's) intensity settings up to round $k$, where $a_j ~(b_j) \in A$ for every $j$. Meanwhile, for simplicity, we assume perfect phase randomization, flawless polarization encoding, and no side channels beyond intensity correlations.

It is noted that the assumptions of transmitters in MDI QKD with intensity correlations are similar to those in the BB84 security framework~\cite{Zapatero2021securityofquantum,PhysRevApplied.18.044069}. To make the security analysis clear, we restate these assumptions used in our security proof. Using Alice as an example, we outline these assumptions while noting that the assumptions for Bob are entirely identical to those for Alice. The following three fundamental assumptions about Alice are used in the security model.

\textbf{Assumption 1:} The Poissonian nature of the photon-number statistics of the source remains uncompromised by the presence of correlations, given the actual intensity $\alpha_k$. Formally, this means that for any given round $k$, and for all $n_k \in \mathbb{N}$,
\begin{align}
    p(n_k|\alpha_k)=\frac{e^{-\alpha_k}\alpha_k^{n_k}}{n_k!}.
    \label{poisson}
\end{align}

\textbf{Assumption 2:} The intensity correlations have a finite range $\xi$, meaning that the actual intensity value $\alpha_k$ is unaffected by the previous settings $a_j$ when $k-j > \xi$.

\textbf{Assumption 3:} We define $\delta$ to represent the relative deviation between intensity setting and actual intensity, i.e.,
\begin{align}
    \delta = |\frac{a_k-\alpha_k}{a_k}|.
    \label{delta}
\end{align}
For round $k$ and any possible record $\vec{a}_k$, one takes the maximum value of $\delta$ as $\delta_{max}$ and define $a_k^\pm = a_k(1\pm\delta_{max})$. Let the function $g_{\vec{a}_k}(\alpha_k)$ represent the average photon number distribution under intensity correlations, with its domain being $[a_k^-,a_k^+]$. Based on the above assumptions, one can derive the photon-number statistics for a given round $k$ and a given record of settings $\vec{a}_k$,
\begin{align}
    p_{n_k}|_{\vec{a}_k}=\int_{a_k^-}^{a_k^+}g_{\vec{a}_k}(\alpha_k)\frac{\text{e}^{-\alpha_k}\alpha_k^{n_k}}{n_k!}d\alpha_k,
    \label{photon-number}
\end{align}
for all $n_k \in \mathbb{N}$.

Although measurement assumptions of receiver are unnecessary in MDI QKD, the definitions of yield and error probability are essential. In this protocol, Alice and Bob post-select the events where they use the same basis and Charlie $\ket{\psi^-}$ or $\ket{\psi^+}$~\cite{PhysRevLett.108.130503}. Thus, based on the MDI QKD protocol and above security assumptions, one can define both the yield and the error probability when the emission photon number of Alice and Bob in round $k$ ($n_k^A,n_k^B \in \mathbb{N}$) is $n$ and $m$, respectively. For example, the yield in Z basis and the error probability in X basis can be expressed respectively as
\begin{widetext}
\begin{align}
    Y_{nm,v_0,\ldots,v_{\xi},w_0,\ldots,w_{\xi}}^{Z,k} &= p(\ket{\psi^-}\cup\ket{\psi^+}| n_{k}^{A} = n, n_{k}^{B} = m, a_{k} = v_{0},\ldots,a_{k-\xi} = v_{\xi},\notag \\
    &~~~~ b_{k} = w_{0},\ldots,b_{k-\xi} = w_{\xi}, x_{k}^{A}=x_{k}^{B} = Z),
    \label{YZ}
\end{align}
\begin{align}
    H_{nm,v_0,\ldots,v_{\xi},w_0,\ldots,w_{\xi}}^{X,k} &= p((\ket{\psi^-},r_k^A=r_k^B)\cup (\ket{\psi^+}, r_k^A\neq r_k^B)|n_{k}^{A} = n, n_{k}^{B} = m, \notag \\ 
    &~~~~a_{k} = v_{0},\ldots, a_{k-\xi} = v_{\xi}, b_{k} = w_{0},\ldots,b_{k-\xi} = w_{\xi}, x_{k}^{A}=x_{k}^{B} = X).
    \label{HX}
\end{align}
\end{widetext}
In the subsequent security proof, our objective is to calculate the lower bound of single-photon yield in $Z$ and $X$ basis, $Y_{11,v_0,\ldots,v_{\xi},w_0,\ldots,w_{\xi}}^{Z,k}$ and $Y_{11,v_0,\ldots,v_{\xi},w_0,\ldots,w_{\xi}}^{X,k}$, and the upper bound of single-photon error probability in $X$ basis, $H_{11,v_0,\ldots,v_{\xi},w_0,\ldots,w_{\xi}}^{X,k}$, thereby estimating the lower bound of the key rate under intensity correlations.

\section{Security proof with intensity correlations}
\label{sec:PROOF}
Inspired by RT~\cite{PereiraSciAdv}, the main idea of our security analysis is to estimate the yield and error probability in the presence of intensity correlations by using the values of yield and error probability from the case without intensity correlations. For the convenience of subsequent descriptions, we refer to the ideal states in the MDI QKD system without intensity correlations as reference states and the states with intensity correlations as actual states. Meanwhile, abbreviating yield and error probability in reference states and actual states as $Y_{nm,ab}^{k,\text{ref}}$, $H_{nm,ab}^{k,\text{ref}}$, $Y_{nm,ab}^{k,\text{act}}$, and $H_{nm,ab}^{k,\text{act}}$, respectively.

One can establish a constraint relationship between reference states and actual states by using CS inequality, which can be expressed as~\cite{PereiraSciAdv,Navarrete_2022,PhysRevApplied.19.044022}
\begin{widetext}
\begin{align}
    &\sqrt{Y_{nm,ab}^{k,\text{ref}} Y_{nm,ab}^{k,\text{act}}} + \sqrt{(1 - Y_{nm,ab}^{k,\text{ref}})(1 - Y_{nm,ab}^{k,\text{act}})} \geq \tau_{ab}^k,\notag \\ 
    &\sqrt{H_{nm,ab}^{k,\text{ref}} H_{nm,ab}^{k,\text{act}}} + \sqrt{(1 - H_{nm,ab}^{k,\text{ref}})(1 - H_{nm,ab}^{k,\text{act}})} \geq \tau_{ab}^k. \label{CS}
\end{align}
\end{widetext}
Here, $\tau_{ab}^k$ represents the squared overlap of yield or error probability between reference states and actual states. Detailed derivations are provided in Refs.~\cite{Zapatero2021securityofquantum,PhysRevApplied.18.044069}. We note that, although the derivations in Refs.~\cite{Zapatero2021securityofquantum,PhysRevApplied.18.044069} are based on the BB84 QKD protocol, they can also be applied to the MDI QKD protocol. The only adjustments needed are to replace the positive operator-valued measurement with a Bell state measurement and to account for two transmitters instead of one. The representation of this parameter in the MDI QKD protocol is
\begin{widetext}
\begin{align}
    &\sqrt{\tau_{ab}^k} = \sum_{ab_{k+1}^{\text{min}\{k+\xi,N\}}}\prod_{i=k+1}^{\text{min}\{k+\xi,N\}}p_{a_i}p_{b_i} \times  \sum_{n=0}^\infty\sum_{m=0}^\infty\sqrt{p_{nm|ab}^\text{act}p_{nm|ab}^\text{ref}}.
    \label{tau}
\end{align}
\end{widetext}
Here, $ab_x^y$ refers to the intensity choices of Alice and Bob from rounds $x$ to $y$, and $p_{nm|ab}^\text{act}/p_{nm|ab}^\text{ref}$ represents the probability that Alice emits $n$ photon(s) and Bob emits $m$ photon(s) with intensity correlations/without intensity correlations. In~\cref{tau}, since there is no intensity correlations in reference states, the value of $p_{nm|ab}^\text{ref}$ can be accurately calculated. Therefore, the lower bound estimation of $\tau_{ab}^k$ depends on the estimation of $p_{nm|ab}^\text{act}$. 

It is worth noting that there are two methods for estimating $p_{nm|ab}^\text{act}$ depending on whether the average photon number distributions under intensity correlations, $g_{\vec{a}_k}(\alpha_k)$ and $g_{\vec{b}_k}(\beta_k)$, are known. When the average photon number distributions are unknown, $p_{nm|ab}^\text{act}$ is estimated pessimistically using the maximum deviations of intensity, $a_k^\pm$ and $b_k^\pm$. When prior information about the average photon number distribution under intensity correlations is available, a more precise estimation of $p_{nm|ab}^\text{act}$ can be made. For example, if the distribution approximately follows a Gaussian distribution, the truncated Gaussian (TG) model can be used to achieve a more accurate estimation of $p_{nm|ab}^\text{act}$~\cite{PhysRevApplied.18.044069}. Further details about these two methods are presented in~\cref{emit_photon_probability}.

For all rounds where Alice's and Bob's intensity choices are $a$ and $b$, the minimum $\tau_{ab}^k$ is found as the lower bound, which is denoted as $\tau_{ab}$. Next, by solving~\cref{CS}, one obtains the range of values for $Y_{nm,ab}^{k,\text{act}}$ and $H_{nm,ab}^{k,\text{act}}$
\begin{align}
    &G_-(Y_{nm,ab}^{k,\text{ref}},\tau_{ab}) \leq Y_{nm,ab}^{k,\text{act}} \leq G_+(Y_{nm,ab}^{k,\text{ref}},\tau_{ab}), \notag \\
    &G_-(H_{nm,ab}^{k,\text{ref}},\tau_{ab}) \leq H_{nm,ab}^{k,\text{act}} \leq G_+(H_{nm,ab}^{k,\text{ref}},\tau_{ab}),
    \label{ACT}
\end{align}
and the range of values for $Y_{nm,ab}^{k,\text{ref}}$ and $H_{nm,ab}^{k,\text{ref}}$
\begin{align}
    &G_-(Y_{nm,ab}^{k,\text{act}},\tau_{ab}) \leq Y_{nm,ab}^{k,\text{ref}} \leq G_+(Y_{nm,ab}^{k,\text{act}},\tau_{ab}), \notag \\
    &G_-(H_{nm,ab}^{k,\text{act}},\tau_{ab}) \leq H_{nm,ab}^{k,\text{ref}} \leq G_+(H_{nm,ab}^{k,\text{act}},\tau_{ab}),
    \label{Ref}
\end{align}
where 
\begin{equation}
    G_{-}(y, z) =
    \begin{cases} 
        g_{-}(y, z) & \text{if } y > 1 - z, \notag\\
        0 & \text{otherwise},
    \end{cases}
\end{equation}

\begin{equation}
    G_{+}(y, z) =
    \begin{cases} 
        g_{+}(y, z) & \text{if } y < z, \\
        1 & \text{otherwise},
    \end{cases}
\end{equation}
with the function $g_{\pm}(y,z)=y+(1-z)(1-2y)\pm 2\sqrt{z}(1-z)y(1-y)$.

As mentioned at the end of~\cref{sec:assumptions}, our goal is to quantitatively provide the lower bound of single-photon yield $Y_{11,ab}^{k,\text{act}}$ and the upper bound of single-photon error probability $H_{11,ab}^{k,\text{act}}$. Based on the convexity of $G_-$ and concavity of $G_+$,~\cref{ACT} can be further scaled as
\begin{align}
    &Y_{11,ab}^{k,\text{act}} \geq G_-(Y_{11,ab}^{k,\text{ref}},\tau_{ab}) \geq G_-(Y_{11,ab}^{\text{ref}},\tau_{ab}) ,\notag \\
    &H_{11,ab}^{k,\text{act}} \leq G_+(H_{11,ab}^{k,\text{ref}},\tau_{ab})\leq G_+(H_{11,ab}^{\text{ref}},\tau_{ab}).  
    \label{Actbound}
\end{align}
Note that $Y_{11,ab}^{\text{ref}}=\sum_{k=1}^N ((a_k=a)\wedge (b_k=b)) Y_{11,ab}^{k,\text{ref}}/\sum_{k=1}^N ((a_k=a)\wedge (b_k=b))$ and $H_{11,ab}^{\text{ref}}=\sum_{k=1}^N ((a_k=a)\wedge (b_k=b)) H_{11,ab}^{k,\text{ref}}/\sum_{k=1}^N ((a_k=a)\wedge (b_k=b))$. In other words, $Y_{11,ab}^{\text{ref}}$ and $H_{11,ab}^{\text{ref}}$ denote the mean yield and the error probability, respectively, under the condition that Alice's intensity setting is $a$ with the emission of a single photon, and concurrently, Bob's intensity setting is $b$ with the emission of a single photon.

To estimate the lower bound of $Y_{11,ab}^{k,\text{act}}$ and the upper bound of $H_{11,ab}^{k,\text{act}}$, one needs to calculate $Y_{11,ab}^{\text{ref}}$ and $H_{11,ab}^{\text{ref}}$ in~\cref{Actbound}. Because there are no intensity correlations in reference states, one can directly solve the mean yield and error probability, using the following equations
\begin{align}
    &Q_{ab}^{\text{ref}} = \sum_{n,m=0} \text{e}^{-(a+b)}\frac{a^n}{n!}\frac{b^m}{m!}Y_{nm}^{\text{ref}}, \notag \\
    &Q_{ab}^{\text{ref}}E_{ab}^{\text{ref}} = \sum_{n,m=0} \text{e}^{-(a+b)}\frac{a^n}{n!}\frac{b^m}{m!}H_{nm}^{\text{ref}}.
    \label{system}
\end{align}
Here, $Q_{ab}^{\text{ref}}$ ($E_{ab}^{\text{ref}}$) represents the gain (quantum bit error rate, QBER) with the Alice's intensity setting $a$ and Bob's intensity setting $b$ in reference states. We omit the subscript $ab$ in $Y_{nm,ab}^{\text{ref}}$ and $H_{nm,ab}^{\text{ref}}$, because in reference states, yield and error probability are independent of intensity setting. 

Substituting three intensity settings $\mu,\nu,\omega$ into~\cref{system}, the lower bound of $Y_{11}^{\text{ref}}$ and the upper bound of $H_{11}^{\text{ref}}$ can be solved by Gaussian elimination, which are expressed as~\cite{PhysRevA.87.012320,Xu_2013}
\begin{widetext}
\begin{align}
    &Y_{11}^{\text{ref}} \geq \frac{(\mu^2-\omega^2)(\mu-\omega)Q^{M1}-(\nu^2-\omega^2)(\nu-\omega)Q^{M2}}{(\mu-\omega)^2(\nu-\omega)^2(\mu-\nu)}, \notag \\
    &H_{11}^{\text{ref}} \leq \frac{\text{e}^{2\nu}Q_{\nu\nu}^{\text{ref}}E_{\nu\nu}^{\text{ref}}+\text{e}^{2\omega}Q_{\omega\omega}^{\text{ref}}E_{\omega\omega}^{\text{ref}}-\text{e}^{\nu+\omega}(Q_{\nu\omega}^{\text{ref}}E_{\nu\omega}^{\text{ref}}+Q_{\omega\nu}^{\text{ref}}E_{\omega\nu}^{\text{ref}})}{(\nu-\omega)^2},
    \label{YLHU}
\end{align}
\end{widetext}
where $Q^{M1}$ and $Q^{M2}$ can be expressed as
\begin{align}
    &Q^{M1}=Q_{\nu\nu}^{\text{ref}}\text{e}^{2\nu}+Q_{\omega\omega}^{\text{ref}}\text{e}^{2\omega}-Q_{\nu\omega}^{\text{ref}}\text{e}^{\nu+\omega}-Q_{\omega\nu}^{\text{ref}}\text{e}^{\omega\nu}, \notag \\
    &Q^{M2}=Q_{\mu\mu}^{\text{ref}}\text{e}^{2\mu}+Q_{\omega\omega}^{\text{ref}}\text{e}^{2\omega}-Q_{\mu\omega}^{\text{ref}}\text{e}^{\mu+\omega}-Q_{\omega\mu}^{\text{ref}}\text{e}^{\omega\mu}.
    \label{QM}
\end{align}
In~\cref{YLHU} and~\cref{QM}, $Q_{ab}^{\text{ref}}$ and $E_{ab}^{\text{ref}}$ ($a,b \in \{\mu,\nu,\omega\}$) are defined in reference and cannot be measured in practice. However, $Q_{ab}^{\text{act}}$ and $E_{ab}^{\text{act}}$ ($a,b \in \{\mu,\nu,\omega\}$), which correspond to actual states, can be directly obtained from the experiment. Therefore, the next objective is to estimate the gain and QBER in reference states using the gain and QBER in actual states. Specifically, one can use~\cref{Ref} and~\cref{system} along with the convexity of the $G_-$ and the concavity of $G_+$ to determine the range of $Q_{ab}^{\text{ref}}$ and $E_{ab}^{\text{ref}}$. The formulas can be expressed as
\begin{gather}
    G_-(Q_{ab}^{\text{act}},\tau_{ab}) \leq Q_{ab}^{\text{ref}} \leq G_+(Q_{ab}^{\text{act}},\tau_{ab}), \notag \\
    G_-(Q_{ab}^{\text{act}}E_{ab}^{\text{act}},\tau_{ab}) \leq Q_{ab}^{\text{ref}}E_{ab}^{\text{Ref}} \leq G_+(Q_{ab}^{\text{act}}E_{ab}^{\text{act}},\tau_{ab}).
    \label{Actboundtwo}
\end{gather}
By applying the constraints on $Q_{ab}^{\text{ref}}$ and $E_{ab}^{\text{ref}}$ from \cref{Actboundtwo}, we can estimate the lower bound for $Y_{11}^{\text{ref}}$ and the upper bound for $H_{11}^{\text{ref}}$ in~\cref{YLHU}, and consequently derive the lower bound for $Y_{11,ab}^{k,\text{act}}$ and $H_{11,ab}^{k,\text{act}}$ in~\cref{Actbound}. Similarly to $\tau_{ab}^k$, for all rounds where Alice's and Bob's intensity choices are $a$ and $b$, find the minimum $Y_{11,ab}^{k,\text{act}}$ as the lower bound, denoted as $Y_{11}^{L}$, and the maximum $H_{11,ab}^{k,\text{act}}$ as the upper bound, denoted as $H_{11}^{U}$. Additionally, by substituting the gain and QBER under specific basis choice (X basis or Z basis) into~\cref{system,YLHU,QM,Actboundtwo}, we can obtain the yield and error probability for this specific basis choice, namely, $Y_{11}^{Z,L}$, $H_{11}^{Z,U}$, $Y_{11}^{X,L}$, and $H_{11}^{X,U}$.

\section{Key rate simulation}
\label{sec:Simulations}
The asymptotic secure key rate of MDI QKD is given by~\cite{PhysRevLett.108.130503,PhysRevA.89.052333}
\begin{align}
    R \geq P_{11}^ZY_{11}^{Z,L}[1-H_2(\frac{H_{11}^{X,U}}{Y_{11}^{X,L}})]-Q_{\mu\mu}^ZfH_2(E_{\mu\mu}^Z).
    \label{keyRate}
\end{align}
Here, $P_{11}^Z$ denotes the probability that Alice and Bob send single-photon states in the $Z$ basis, i.e., $P_{11}^Z = q_{Z^A}q_{Z^B}\sum_{a \in A}\sum_{b \in A}p_{11|ab}^{act}$. $f$ is the bidirectional error correction efficiency, and $H_2(x) = -x \log_2(x) - (1 - x) \log_2(1 - x)$ is the binary Shannon information entropy. Other parameters have been defined in~\cref{sec:assumptions,sec:PROOF}.

In the simulation, $Q_{ab}$ and $E_{ab}$ depend on Alice and Bob's intensity settings $a$ and $b$, the total transmittance of the channel and detector $\eta$, the probability of background counts $Y_0$, and misalignment errors $e_d$.~\cref{QEcalculations} provides the calculations of $Q_{ab}$ and $E_{ab}$ using these parameters. Consistent with Refs.~\cite{Zapatero2021securityofquantum,PhysRevApplied.18.044069}, since we are considering the asymptotic regime, the key rate does not depend on the probabilities of decoy settings and basis choices. Therefore, we setting $p_{\mu} \approx 1$ and $q_Z \approx 1$ to maximizes key rate similar to Ref.~\cite{PhysRevApplied.18.044069}. The other simulation parameters are utilized from Ref.~\cite{PhysRevX.10.031030} as follows. The bidirectional error correction efficiency to $f=1.16$, the probability of background counts $Y_0 = 4\times10^{-8}$, and misalignment errors $e_d=0.0108$. For the total transmittance of the channel and detector, $\eta = \eta_d10^{-\alpha L/10}$, where $\eta_d$ is the detection efficiency of the detector, $\alpha$ is the attenuation coefficient of the channel, and $L$ is the distance from Alice (Bob) to Charlie. In the simulation, we set $\eta_d=0.53$ and $\alpha=0.2$ dB/km.

\begin{figure}[h]
    \centering
    \includegraphics[width=\linewidth]{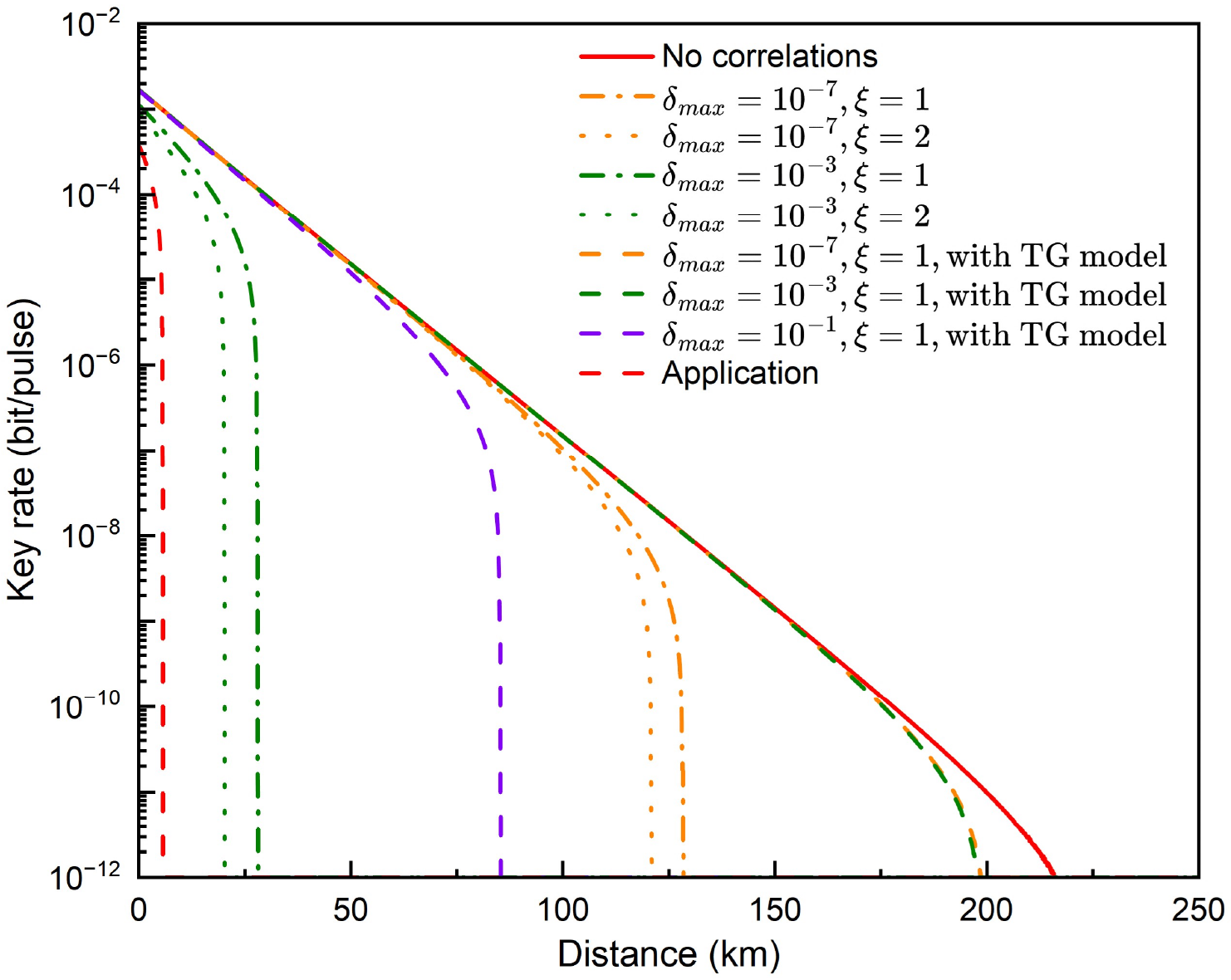}
    \caption{{\bfseries The key rate for symmetric MDI QKD with intensity correlations.} \justifying{We fix Alice and Bob's intensities to $\mu=0.207,~\nu=0.035, ~\text{and}~\omega=10^{-4}$. Other parameters related to the channel and detectors are provided in~\cref{sec:Simulations}. The solid red line represents the key rate without intensity correlations. The dotted and dot-dashed lines indicate the key rate when the average photon number distributions under intensity correlations cannot be determined. Specifically, the orange and green dot-dashed lines show the key rate for $\delta_{max} \in \{10^{-7},10^{-3}\}$ and $\xi =1 $, while the orange and green dotted lines represent the key rate for $\delta_{max} \in \{10^{-7},10^{-3}\}$ and $\xi = 2$. The dashed lines show the key rate curve when the average photon number distributions under intensity correlations follow Gaussian distributions. For clarity, only the key rate for $\xi = 1$ is plotted in this case. The orange, green, and purple dashed lines correspond to the key rates for $\delta_{max}=10^{-7},10^{-3},10^{-1}$ with TG model, respectively. The red dashed line represents the simulated key rate when applying our theoretical model to specific experiments presented in~\cref{sec:Application}.}}
    \label{Fig:keyRate}
    \end{figure}

If prior information about $g_{\vec{a}_k}(\alpha_k)$ is unknown, the key rate depends on Alice's (Bob's) intensity settings, the maximum relative deviation $\delta_{max}$, and the intensity correlations range $\xi$. Additionally, if the information about $g_{\vec{a}_k}(\alpha_k)$ is available, such as when the average photon number approximately follows a Gaussian distribution, one can  use TG model to more accurately estimate the key rate. We fix Alice's and Bob's intensity settings, and the simulated key rate is depicted in~\cref{Fig:keyRate}. When the average photon number distribution under intensity correlations cannot be obtained, our simulations indicate that the key rate decreases as the maximum relative deviation $\delta_{max}$ or the correlation length $\xi$ increases. Specifically, when $\delta_{max}=10^{-7}$, the maximum key transmission distance is reduced to approximately half of that without intensity correlations. As $\delta_{max}$ increases to $10^{-3}$, the maximum key transmission distance drops to around $25$ km. From~\cref{Fig:keyRate}, it can be observed that while the use of the TG model significantly improves the key rate, the maximum transmission distances for $\delta_{max}=10^{-7}$ and $10^{-3}$ are comparable to those without intensity correlations. However, when $\delta_{max}$ continues to increase to $10^{-1}$, the maximum key transmission distance is reduced to approximately one-third of the distance without intensity correlations. Our simulation results show that deviations caused by intensity correlations have a significant impact on the secret key rate.

\section{Application}
\label{sec:Application}

\begin{figure*}[htb]
    \centering
    \subfloat[The average photon number distribution for $SS$.]{\label{Fig:SS-distribution}\includegraphics[width=0.45\textwidth]{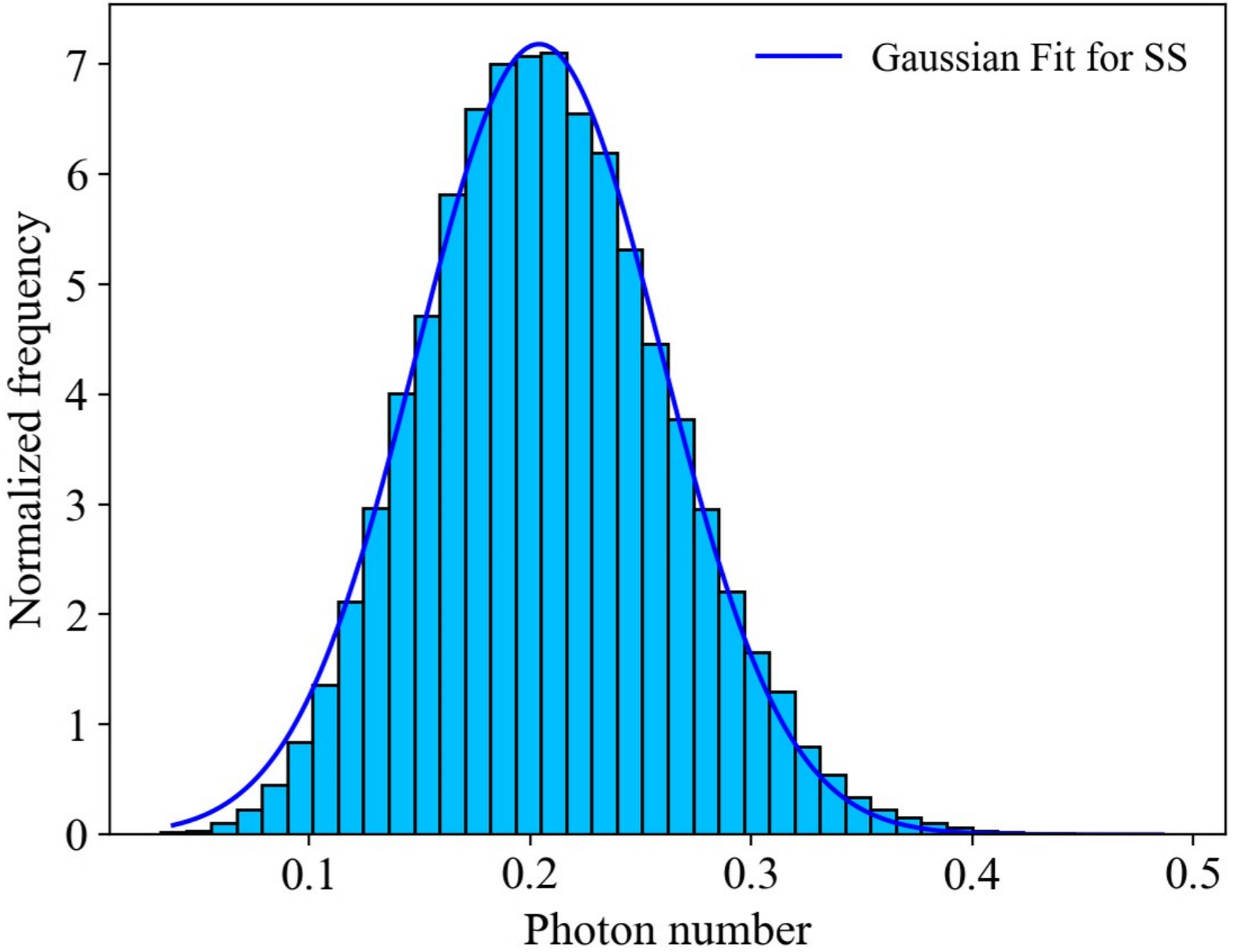}}
    \hfill
    \subfloat[The average photon number distribution for $S$.]{\label{Fig:S-distribution}\includegraphics[width=0.45\textwidth]{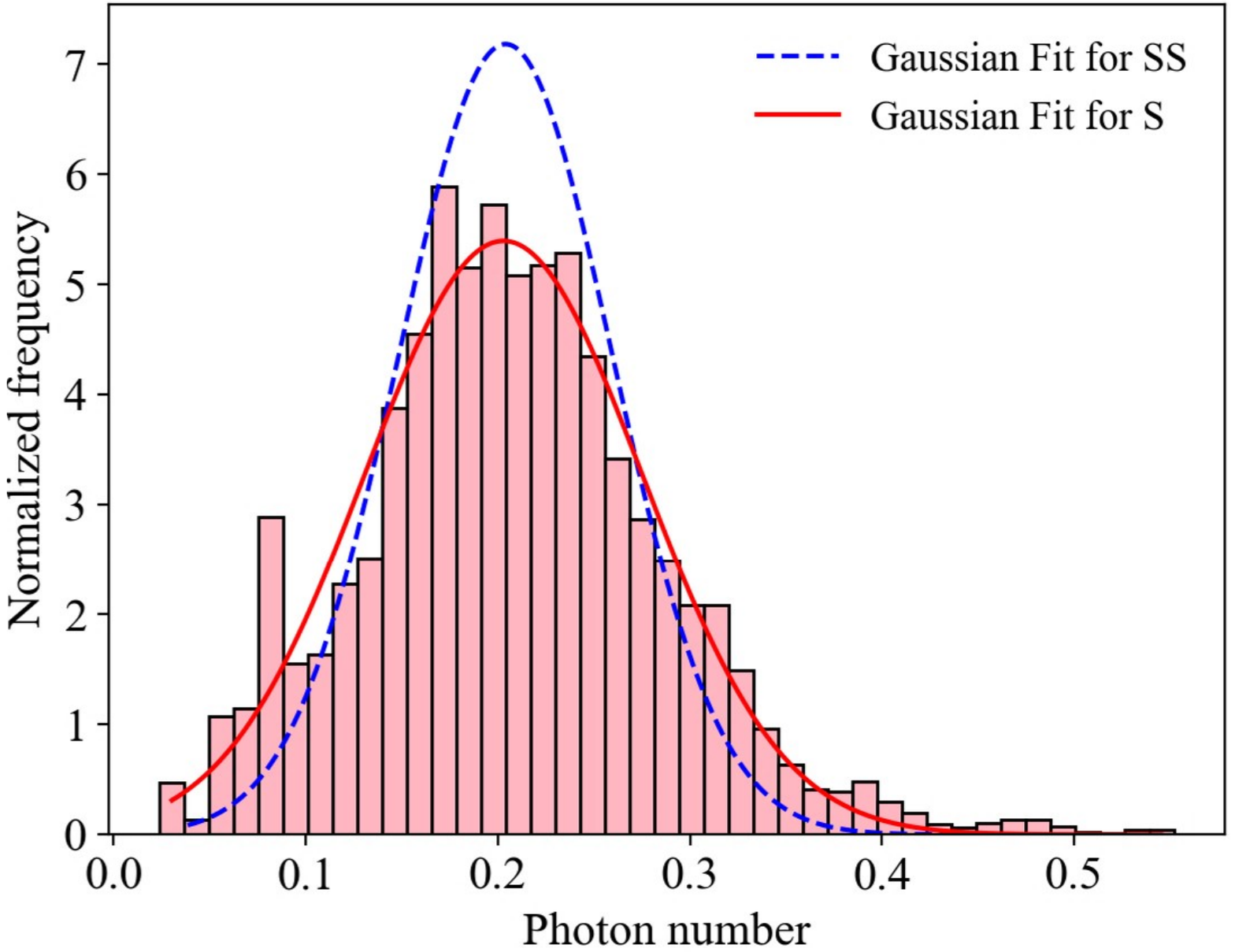}}
    \caption{ Intensity distribution of signal state measured from the MDI QKD system under test. (a) The distribution of $SS$ follows Gaussian distribution with an mean value of 0.204 and a standard deviation of 0.056. (b) Weighted statistics of $VS$, $D_1S$, $D_2S$, and $SS$ give the distribution of $S$, which roughly conforms to the Gaussian distribution with a mean of 0.203 and a standard deviation of 0.072. The blue and red curves respectively represent the Gaussian fitting results of SS and S.}
\end{figure*}

Considering the correlation features of a MDI QKD system in practice, the theoretical model can be applied to the MDI QKD system and illustrates its actual secure key rate. The MDI QKD system we take as an example contains four intensities, $V$, $D_1$, $D_2$ and $S$, and sends a random number sequence repeatedly for $1.52 \times 10^6$ times~\cite{Xing:24}. The detection results from the single-photon detector~(SPD) are obtained in the experiment. Without prejudice to generality, we take Alice's side with its intensity correlations $\xi=1$ as an example to demonstrate the impact on the secure key rate, which is simple and practical for a MDI QKD system. There are the following three steps to extract the correlation parameters from the measurement result. 

First, the intensity patterns are divided into different groups and the click rate distributions of the groups are calculated. Since $\xi=1$, the correlated intensity patterns include two intensities, $a_1$ and $a_2$, containing 16 groups ($[a_1a_2]$, $a_1,a_2=V, D_1, D_2$ and $S$). According to the detection result, the click rate distribution of each group can be calculated by the correlation characterization via SPD method in the MDI QKD system as presented in Ref.~\cite{Xing:24}. Without affecting generality, here we take the groups of the signal state ($a_2=S$) as an example. We can calculate the click rate distributions of $VS$, $D_1S$, $D_2S$, and $SS$, whose average click rates are \num{1.2e-4}, \num{1.3e-4}, \num{1.7e-4} and \num{1.8e-4}, respectively.

Second, the click rate distributions of groups are converted into their average photon number distributions. Since we consider the click rate linearly represent the photon number~\cite{Xing:24}, it is feasible to obtain the average photon number distribution according to the click rate distribution. For example, the average photon number distribution of the $SS$ group can be obtained from the click rate distribution as shown in Fig.~\ref{Fig:SS-distribution}, which approximately follows a Gaussian distribution. Similarly, the average photon number distributions of other groups can be calculated from their click rate distributions.

Third, the actual average photon number distribution of the signal state is weighted statistics obtained from the distributions of $VS, D_1S, D_2S$ and $SS$. The weighted statistics contains two part of weights, the sending weight from the source unit and the detection weight from the measurement unit. On one hand, since the sending times of different groups vary, the click rates from the groups have different sending weights $w_{send}$, which equals to the normalized ratio of sending numbers. On the other hand, the measurement samples for the groups are different and the detection weight for $a_1S$, $w_{det}$, equals to its ratio of the measurement samples. Eventually, the weight of each group is the product of the sending weight and the detection weight, i.e. $w=w_{send} \times w_{det}$. According to the data from the MDI QKD system, we have $w_{VS}: w_{D_1S}: w_{D_2S}: w_{SS}=  0.061: 0.253: 0.083: 0.603$, finally obtaining the distribution of the actual intensity $S$, shown in Fig.~\ref{Fig:S-distribution}.

It is worth noting that, according to the security proof of decoy-state protocol, the key rate obtained using the signal state, weak decoy state, and vacuum state can approximate the key rate obtained with an infinite number of decoy states~\cite{PhysRevA.72.012326}. Therefore, to simplify calculations while maintaining accuracy, we choose intensities of signal state $S$, weak decoy state $D_1$, and vacuum state $V$ in our simulation. Next, taking the signal state $S$ as an example, we analyze the relative deviation $\delta$ under intensity correlations.

As shown in Fig.~\ref{Fig:SS-distribution}, the intensity distribution of $SS$ can be fitted to a Gaussian distribution $N(0.204, 0.056^2)$, which attributes to random fluctuations caused by the SPD and other system fluctuations. Additionally, as shown in~\cref{Fig:S-distribution}, the intensity distribution of $S$ can be fitted to a Gaussian distribution $N(0.203, 0.072^2)$, which is considered as the result of both random fluctuations and intensity correlations. Assuming that the effects of random fluctuations and intensity correlations on the emission intensity are independent, the deviation caused solely by intensity correlations follows Gaussian distribution $N(-0.001, 0.045^2)$. This shows that intensity correlations can lead to a decrease in the average intensity by 0.001 and an increase in fluctuations with a standard deviation of 0.045. Using the $3\sigma$ rule of the Gaussian distribution, we can calculate the range of intensity fluctuations caused by intensity correlations as $[-0.136,0.134]$. Since we consider the intensity of the $SS$ as the intensity without intensity correlations, therefore, intensity setting $a = 0.204$ and actual intensity $\alpha \in [a-0.136,a+0.134]$. Therefore, based on Assumption 3 in~\cref{sec:assumptions}, the relative deviation can be calculated as $\delta = \frac{|a-\alpha|}{a} \approx 0.666$. Similar to the simulation in~\cref{sec:Simulations}, we assume $p_{\mu} \approx 1$. Therefore, estimating the fluctuation range of the signal state under intensity correlations is sufficient to calculate the secure key rate. When the average photon number distributions are unknown, no positive key rate is generated, whereas, when the average photon number distributions follow Gaussian distribution, and the TG model is used, the key rate is depicted by the red dashed line in~\cref{Fig:keyRate}, indicating that the key can only be transmitted over approximately 5 km.

From the security analysis and key-rate simulation, it is clear that the intensity correlations compromise the security of the tested MDI QKD system. Furthermore, the security boundary under intensity correlations can be explored by this security model. Taking the tested MDI QKD system as an example, in our theoretical model, when the average photon number distributions are unknown and the maximum relative deviation due to intensity correlations is $10^{-2}$, the key can only be transmitted over 1 km. If the intensity correlations are larger than this boundary value, the MDI QKD system cannot produce secret key. When the average photon number distributions are known and the maximum relative deviation due to intensity correlations reaches 0.8, the system is unable to generate any positive secure keys.

\section{Conclusion}
\label{sec:Conclusion}

This paper proposes a theoretical model that quantifies the impact of intensity correlations on the security of MDI QKD system. Based on RT~\cite{PereiraSciAdv} and the previous frameworks of intensity correlations~\cite{Zapatero2021securityofquantum,PhysRevApplied.18.044069}, our theoretical model uses the CS inequality to establish a constraint relationship between reference states (without intensity correlations) and actual states (with intensity correlations). We simulate the key rates generated by the MDI QKD system under intensity correlations. Furthermore, we apply the theoretical model to a practical MDI QKD system to estimate its secret key rate. Our theoretical analysis indicates that intensity correlations significantly compromise the practical security of MDI QKD systems. This security model also provides an approach to evaluate the boundary conditions for key generation in MDI QKD systems. It is recommended that the intensity fluctuations due to intensity correlations should be kept low enough to maintain a positive key rate.

The theoretical model extends the security analysis of intensity correlations to the MDI QKD protocol, thereby enriching the security framework of MDI QKD systems. Additionally, this paper applies the theoretical model to a practical MDI QKD system to show its feasibility and analyze the security performance of this system. Simulations of the key rate show that, compared to the BB84 QKD system, the security of the MDI QKD system is more susceptible to intensity correlations. Therefore, intensity correlations is necessary to be considered in a MDI QKD system. Moreover, the boundary conditions for key generation in MDI QKD systems proposed by this theoretical model may offer new insights for future research on the stability of the source in MDI QKD systems. In summary, this work proposes the methodology to analyse the intensity correlations in a MDI QKD system, as a further step to approach the practical security in high-speed MDI QKD implementation.

\section*{Funding.} Innovation Program for Quantum Science and Technology (2021ZD0300704); National Natural Science Foundation of China (Grant No.\ 62371459); National Key Research and Development Program of China (2019QY0702).

\section*{Acknowledgement.} We thank Jialei Su and Qingquan Peng for helpful discussions, and thank Wei Li, Likang Zhang, and Feihu Xu for data supporting.

\section*{Disclosures.}The authors declare no conflicts of interest.
\appendix

\newpage
\section*{Appendix}

    \section{Probability estimation of photon numbers emitted by the transmitter with intensity correlations} 
    \label{emit_photon_probability}

    In this appendix, we present two methods for estimating the number of photons emitted by the transmitter with intensity correlations. These methods have different conditions of applicability. The first method applies when the maximum relative deviation $\delta_{max}$, as assumed in~\cref{sec:assumptions}, is known, but the distributions of average photon numbers $g_{\vec{a}_k}(\alpha_k)$ and $g_{\vec{b}_k}(\beta_k)$ are unknown. In this case, we use $a_k^\pm$ and $b_k^\pm$ to determine the range of values for $p_{nm|ab}^\text{act}$ in~\cref{tau} of the main text. Based on the different values of $n$ and $m$, the range of values for $p_{nm|ab}^\text{act}$ is as follows
\begin{widetext}
\begin{equation}
p_{nm|ab}^\text{act} \in
\begin{cases}
[P(n|a^+)P(m|b^+),P(n|a^-)P(m|b^-)] & \text{if }n=0,m=0. \\
[P(n|a^-)P(m|b^+),P(n|a^+)P(m|b^-)] & \text{if }n>0,m=0. \\
[P(n|a^+)P(m|b^-),P(n|a^-)P(m|b^+)] & \text{if }n=0,m>0. \\
[P(n|a^-)P(m|b^-),P(n|a^+)P(m|b^+)] & \text{if }n>0,m>0. \\
\end{cases}
\label{first_method}
\end{equation}
\end{widetext}
Here, $P(n|\lambda)$ represents the probability that the number of photons in a pulse is $n$, given that the weak coherent source intensity is selected as $\lambda$. This probability follows a Poisson distribution, expressed as $P(n|\lambda)=\frac{\lambda^n}{n!}\text{e}^{-\lambda}$. It is important to note that in~\cref{first_method} and the subsequent equations in this appendix, we have omitted the subscript $k$ for consistency with~\cref{tau}.

The second method is applicable when prior information about the average photon number distributions $g_{\vec{a}_k}(\alpha_k)$ and $g_{\vec{b}_k}(\beta_k)$ under intensity correlations are available. In this case, we can more precisely determine the value of $p_{nm|ab}^\text{act}$. According to~\cref{photon-number}, and given the independence between Alice and Bob, we have
\begin{align}
    p_{nm|ab}^\text{act}=\int_{a^-}^{a^+}g_{\vec{a}}(\alpha)\frac{\text{e}^{-\alpha}\alpha^{n}}{n!}d\alpha\int_{b^-}^{b^+}g_{\vec{b}}(\beta)\frac{\text{e}^{-\beta}\beta^{m}}{m!}d\beta,
    \label{photon-number-Appendix}
\end{align}
For example, when the average photon number distributions are approximately Gaussian distributions, Ref.~\cite{PhysRevApplied.18.044069} proposes a truncated Gaussian (TG) model to accurately estimate $p_{nm|ab}^\text{act}$. Specifically, Ref.~\cite{PhysRevApplied.18.044069} indicates that within a fixed finite interval $[\lambda,\Lambda]$, truncated Gaussian distribution can be expressed as
\begin{equation}
g_{\vec{a}}(\gamma, \sigma, \lambda, \Lambda; \alpha) = 
\begin{cases} 
0 & \text{if } \alpha \leq \lambda, \\ 
\frac{\phi(\gamma, \sigma^2; \alpha)}{\Phi(\gamma, \sigma^2; \Lambda) - \Phi(\gamma, \sigma^2; \lambda)} & \text{if } \lambda < \alpha < \Lambda, \\ 
0 & \text{if } \Lambda \leq \alpha,
\label{TGdistribution}
\end{cases}
\end{equation}
where
\begin{align}
\phi(\gamma, \sigma^2; x) &= \frac{1}{\sigma \sqrt{2\pi}} \text{e}^{-\frac{(x-\gamma)^2}{2\sigma^2}}, \notag \\
\Phi(\gamma, \sigma^2; x) &= \int_{-\infty}^{x} \frac{1}{\sigma \sqrt{2\pi}} \text{e}^{-\frac{(t-\gamma)^2}{2\sigma^2}} \text{d}t.
\label{Phiphi}
\end{align}
$\gamma$ and $\sigma^2$ are the mean and variance of the average photon number distribution. By substituting the function $g$ from~\cref{TGdistribution} into~\cref{photon-number-Appendix}, one can estimate $p_{nm|ab}^\text{act}$ under TG model.

By substituting the estimated $p_{nm|ab}^\text{act}$ from~\cref{first_method} or~\cref{photon-number-Appendix} into~\cref{tau}, one can determine the squared overlap $\tau_{ab}^k$. It is important to note that the accuracy of the estimation of $p_{nm|ab}^\text{act}$ directly influences the tightness of the bounds on $\tau_{ab}^k$, which in turn significantly impacts the estimation of the key rate discussed later in main text.
    
    \section{Calculations of gain and QBER in simulation} 
    \label{QEcalculations}
    Based on Ref.~\cite{Xu_2013}, we performed the simulation calculations of the gain and QBER by making prior assumptions about the source, channel, and detector. For simplicity, unlike Ref.~\cite{Xu_2013}, we consider a symmetric MDI QKD protocol where Alice and Bob have the same channel transmittance. Next, we provide the calculation steps for $Q_{ab}$ and $E_{ab}$ using the Z basis as an example. The process for the X basis is similar.

    According to the MDI QKD protocol and the definitions of parameters $Q_{ab}$ and $E_{ab}$, we have
\begin{align}
Q_{ab} = \frac{Q_{ab}^{HH}+Q_{ab}^{HV}}{2} \notag \\
E_{ab} = \frac{Q_{ab}^{HH}}{Q_{ab}^{HH}+Q_{ab}^{HV}}
\label{QEfinnal}
\end{align}
    The superscript, $HH$ or $HV$ denote Alice’s and Bob’s encoding modes. Also, $Q_{ab}^{HH}$ and $Q_{ab}^{HV}$ is given by
\begin{align}
Q_{ab}^{HH} = Q_{ab}^{HH,\psi^+}+Q_{ab}^{HH,\psi^-} \notag \\
Q_{ab}^{HV} = Q_{ab}^{HV,\psi^+}+Q_{ab}^{HV,\psi^-}
\label{QHHV}
\end{align}
Here, the superscripts $\psi^+$ and $\psi^-$ denote the Bell states post-selected by Alice and Bob.

For the calculation on the RHS of~\cref{QHHV}, the conclusion from Ref.~\cite{Xu_2013} is as follows
\begin{widetext}
\begin{align}
Q_{ab}^{HH,\psi^+} &= 2\text{e}^{-\frac{\gamma}{2}}(1-Y_0)^2[I_0(\beta)+(1-Y_0)^2\text{e}^{-\frac{\gamma}{2}} \notag \\
&~~~~-(1-Y_0)\text{e}^{-\frac{\gamma(1-e_d)}{2}}I_0(e_d\beta)-(1-Y_0)\text{e}^{-\frac{\gamma e_d}{2}}I_0(\beta-e_d\beta)], \notag \\
Q_{ab}^{HH,\psi^-} &= 2\text{e}^{-\frac{\gamma}{2}}(1-Y_0)^2[I_0(\beta-2\beta e_d)+(1-Y_0)^2\text{e}^{-\frac{\gamma}{2}} \notag \\
&~~~~-(1-Y_0)\text{e}^{-\frac{\gamma(1-e_d)}{2}}I_0(e_d\beta)-(1-Y_0)\text{e}^{-\frac{\gamma e_d}{2}}I_0(\beta-e_d\beta)], \notag \\
Q_{ab}^{HV,\psi^+} &= 2\text{e}^{-\frac{\gamma}{2}}(1-Y_0)^2[I_0(2\lambda)+(1-Y_0)^2\text{e}^{-\frac{\gamma}{2}} \notag \\
&~~~~-(1-Y_0)\text{e}^{-\frac{\zeta}{2}}I_0(\lambda)-(1-Y_0)\text{e}^{-\frac{\gamma -\zeta}{2}}I_0(\lambda)], \notag \\
Q_{ab}^{HV,\psi^-} &= 2\text{e}^{-\frac{\gamma}{2}}(1-Y_0)^2[1+(1-Y_0)^2\text{e}^{-\frac{\gamma}{2}} \notag \\
&~~~~-(1-Y_0)\text{e}^{-\frac{\zeta}{2}}I_0(\lambda)-(1-Y_0)\text{e}^{-\frac{\gamma -\zeta}{2}}I_0(\lambda)]. 
\label{psi+-}
\end{align}
\end{widetext}
The polynomial expansions of functions $I_0(x)$ and $\text{e}^x$ are given by
$I_0(x) = 1+\frac{x^2}{4} + O(x^4)$, $\text{e}^x = 1+x+\frac{x^2}{2}+O(x^3)$. The parameters $\gamma$, $\beta$, $\lambda$, and $\zeta$ are coefficients related to Alice's and Bob's intensity settings, channel modeling, and detection efficiency. The specific calculations are as follows
\begin{align}
\gamma_{Alice} &= \sqrt{a\eta}, \notag \\ 
\gamma_{Bob} &= \sqrt{b\eta}, \notag \\ 
\beta &= \gamma_{Alice}\gamma_{Bob}, \notag \\ 
\gamma &= \gamma_{Alice}^2+\gamma_{Bob}^2, \notag \\ 
\lambda &= \gamma_{Alice}\gamma_{Bob}\sqrt{e_d(1-e_d)}, \notag \\ 
\zeta &= \gamma_{Alice}^2+e_d(\gamma_{Bob}^2-\gamma_{Alice}^2).
\label{coefficients}
\end{align}
By substituting the simulation parameters from the main text into~\cref{psi+-,coefficients}, the values of $Q_{ab}^{HH}$ and $Q_{ab}^{HV}$ in~\cref{QHHV} can be obtained. This, in turn, allows for the calculation of gain and QBER in~\cref{QEfinnal}.


%

\end{document}